\catcode`\@=11
\newif\if@fewtab\@fewtabtrue

{\count255=\time\divide\count255 by 60
\xdef\hourmin{\number\count255}
\multiply\count255 by-60\advance\count255 by\time
\xdef\hourmin{\hourmin:\ifnum\count255<10 0\fi\the\count255}}
\def\ps@draft{\let\@mkboth\@gobbletwo
    \def\@oddhead{}
    \def\@oddfoot
       {\hbox to 7 cm{$\scriptstyle Draft\ version:\ \draftdate$
       \hfil}\hskip -7cm\hfil\rm\thepage \hfil}
    \def\@evenhead{}\let\@evenfoot\@oddfoot}

\def\ceqno{\global\@fewtabfalse
    \ifcase\@eqcnt \def\@tempa{& & &}\or \def\@tempa{& &}
      \or \def\@tempa{&}
      \or\def\@tempa{}\fi\@tempa
{\rm(\theequation)}}

\def\aeqno#1{\global\@fewtabfalse
    \ifcase\@eqcnt \def\@tempa{& & &}\or \def\@tempa{& &}
      \or \def\@tempa{&}
      \or\def\@tempa{}\fi\@tempa
{\rm(\theequation,#1)}}

\def\label#1{\ifnum\draftcontrol=1
 \global\def\draftnote{$\scriptstyle #1$}\fi
 \@bsphack\if@filesw {\let\thepage\relax
   \def\protect{\noexpand\noexpand\noexpand}%
\xdef\@gtempa{\write\@auxout{\string
      \newlabel{#1}{{\@currentlabel}{\thepage}}}}}\@gtempa
   \if@nobreak \ifvmode\nobreak\fi\fi\fi
  \@esphack}

\def\alabel#1#2{\label{#1}\global\@fewtabfalse
    \ifcase\@eqcnt \def\@tempa{& & &}\or \def\@tempa{& &}
      \or \def\@tempa{&}
      \or\def\@tempa{}\fi\@tempa
{\hbox to 3cm{\phantom{\rm(\theequation,#2)}
\draftnote \hfil}\hskip -3cm {\rm(\theequation,#2)}}}

\def\clabel#1{\label{#1}\global\@fewtabfalse
    \ifcase\@eqcnt \def\@tempa{& & &}\or \def\@tempa{& &}
      \or \def\@tempa{&}
      \or\def\@tempa{}\fi\@tempa
{\hbox to 3cm{\phantom{\rm(\theequation)}
\draftnote \hfil}\hskip -3cm{\rm(\theequation)}}}

\def\eqnarray{\def\draftnote{{}}\global\@fewtabtrue
\stepcounter{equation}\let\@currentlabel=\theequation
\global\@eqnswtrue
\global\@eqcnt\z@\tabskip\@centering\let\\=\@eqncr
$$\halign to \displaywidth\bgroup\@eqnsel\hskip\@centering\@eqcnt\z@
  $\displaystyle\tabskip\z@{##}$&\global\@eqcnt\@ne
  \hskip 1\arraycolsep \hfil${##}$\hfil
  &\global\@eqcnt\tw@ \hskip 1\arraycolsep
$\displaystyle\tabskip\z@{##}$
\hfil  \tabskip\@centering&\global\@eqcnt\thr@@\llap{##}\tabskip\z@
\cr}

\def\endeqnarray{\@@eqncr\egroup
      \global\advance\c@equation\m@ne$$\global\@ignoretrue}

\def\@eqnnum{\hbox to 3cm{\phantom{\rm(\theequation)} \draftnote
                         \hfil}\hskip -3cm {\rm(\theequation)}}

\def\@@eqncr{\let\@tempa\relax
    \ifcase\@eqcnt \def\@tempa{& & &}\or \def\@tempa{& &}
      \or \def\@tempa{&}
      \or\def\@tempa{}
\fi\@tempa
\if@eqnsw
\if@fewtab\@eqnnum\fi
\stepcounter{equation}\fi\global
\@eqnswtrue\global\@eqcnt\z@\global\@fewtabtrue\cr}

\def\draftcite#1{\ifnum\draftcontrol=1#1\else{}\fi}

\def\@lbibitem[#1]#2{\item{}\hskip -3cm \hbox to 2cm
{\hfil$\scriptstyle\draftcite{#2}$}\hskip
1cm[\@biblabel{#1}]\if@filesw
     {\def\protect##1{\string ##1\space}\immediate
      \write\@auxout{\string\bibcite{#2}{#1}}}\fi\ignorespaces}

\def\@bibitem#1{\item\hskip -3cm \hbox to 2cm
{\hfil $\scriptstyle\draftcite{#1}$}\hskip 1cm
\if@filesw \immediate\write\@auxout
       {\string\bibcite{#1}{\the\value{\@listctr}}}\fi\ignorespaces}

\font\tendl=msbm10  scaled \magstep1%double line
\font\sevendl=msbm7 scaled \magstep1
\font\fivedl=msbm5 scaled \magstep1
\font\tengl=eufm10  scaled \magstep1% gothic letters
\font\sevengl=eufm7 scaled \magstep1
\font\fivegl=eufm5 scaled \magstep1

\newfam\dlfam  % \dl is double line
\textfont\dlfam=\tendl \scriptfont\dlfam=\sevendl
\scriptscriptfont\dlfam=\fivedl
\newfam\glfam  % \gl is gothic letters
\textfont\glfam=\tengl \scriptfont\glfam=\sevengl
\scriptscriptfont\glfam=\fivegl

\def\draftdate{\number\month/\number\day/\number\year\ \ \ \hourmin }

\global\def\draftcontrol{0}
\catcode`\@=12
\def\tilde{\widetilde}

\documentstyle[12pt]{article}

\renewcommand{\theequation}{\arabic{equation}}
\def\theequation{{\arabic{equation}}}

\newcommand{\be}{\begin{eqnarray}}
\newcommand{\en}{\end{eqnarray}\vs 0.5 cm}

\newcommand{\vs}{\vskip}
\newcommand{\VS}{\vspace}
\newcommand{\hs}{\hspace}

\newcommand{\NR}{{{\bf R}}}%letra doble raya en modo matematico
\newcommand{\NZ}{{{\bf Z}}}%letra doble raya en modo matematico
\newcommand{\NX}{{{\bf X}}}
\newcommand{\NY}{{{\bf Y}}}

\newcommand{\Nx}{{{\bf x}}}

\newcommand{\Nv}{{{\bf v}}}

\newcommand{\Nk}{{{\bf k}}}

\newcommand{\qq}{\begin{eqnarray}}

\newcommand{\da}{\partial}
\newcommand{\ee}{{\rm e}}

\newcommand{\qqq}{\end{eqnarray}}

\newcommand{\CC}{{\cal C}}
\newcommand{\CD}{{\cal D}}

\newcommand{\CL}{{\cal L}}
\newcommand{\CM}{{\cal M}}

\newcommand{\CO}{{\cal O}}

\newcommand{\CT}{{\cal T}}

\newcommand{\ha}{{1\over 2}}

\newcommand{\s}{\hspace{0.05cm}}
\newcommand{\m}{\hspace{0.025cm}}

\newcommand{\hf}{{_1\over^2}}

\begin{document}

\begin{center}
\

\vs 1cm
{\Large{\bf{Anomalous Scaling of the Passive Scalar}}}

\vs 1cm
{\large{Krzysztof Gaw\c{e}dzki}}
\vs 0.2cm
I.H.E.S., C.N.R.S.,
F-91440  Bures-sur-Yvette, France
\vs 0.5cm
{\large{Antti Kupiainen}}\footnote{Partially supported by
NSF grant DMS-9205296 and EC grant CHRX-CT93-0411}
\vs 0.2cm
Mathematics Department, Helsinki University,

PO Box 4, 00014 Helsinki, Finland
\end{center}
\date{ }
%\maketitle

%%% for draft versions, suppress in definitive version:
%\draft
%%
%%% suppress in definite version:
%\vskip 0.3cm
%\begin{center}
%(preliminary version)
%\end{center}
%%%
\vskip 1.3 cm

\begin{abstract}
\vskip 0.3cm

\noindent We establish anomalous inertial range scaling
of structure functions for a model of  advection of a
passive scalar by a random velocity field.
The velocity statistics is taken gaussian with
decorrelation in time and velocity differences scaling as
$|x|^{\kappa/2}$ in space, with $0\leq\kappa < 2$.
The scalar is driven by a gaussian forcing acting on spatial
scale $L$ and decorrelated in time. The
structure functions for the scalar are well defined
as the diffusivity is taken to zero and acquire anomalous
scaling behavior for large pumping scales $L$.
The anomalous exponent is calculated explicitly for
the $4^{\m\rm th}$ structure function and for small $\kappa$
and it differs from previous predictions. For all but the
second structure functions the anomalous exponents are
nonvanishing.
\end{abstract}
\vs 1.6cm

In 1941 A.N. Kolmogorov argued that in fully developed
turbulence a range of spatial scales exists where
the velocity structure functions acquire a form
independent of the IR and UV cutoffs provided
by the scale of energy pumping and dissipation
respectively. Ever since  a debate
has been going on  as to whether there are corrections
to the scaling exponents predicted by Kolmogorov and
whether such corrections depend on the dissipation
or the pumping scale or both.
This question being still quite open for Navier-Stokes
turbulence both experimentally and theoretically,
it is useful to consider it in the context of
simpler models that are nevertheless expected
to display phenomena similar to the Navier-Stokes
equations.

In this letter we consider one such model that has attracted
much attention recently \cite{Kr94, LPF, Majda, Russ, Proc,
Kra}, namely
that of passive advection in a random velocity
field \s$\Nv(t,\Nx)\s$ of a scalar quantity \s$T\s$
whose density \s$T(t,\Nx)\s$ satisfies the equation
\qq
\da_t\m T\s+\s(\Nv\cdot\nabla)\m T\s-\s\nu\Delta T\s=\s f\ ,
\label{PS}
\qqq
where  \s$\nu\s$ denotes the molecular diffusivity
of the scalar \s$T\s$
and \s$f(t,\Nx)\s$ is an external source driving the system.

We take \s$\Nv(t,\Nx)\s$ and  \s$f(t,\Nx)\s$ to be
mutually independent Gaussian
random fields with zero mean and covariances
\qq
< v^i(t,\Nx)\m\s v^j(t',\Nx')>
\s=\s\delta(t-t')\s\s D^{ij}(\Nx-\Nx')\ ,
\label{42}
\qqq
\qq
< f(t,\Nx)\s\m f(t',\Nx')>
\s=\s\delta(t-t')\ \CC({_{\Nx-\Nx'}\over^L})
\m\equiv\s\delta(t-t')\ \CC_L(\Nx-\Nx')\ .
\label{force}
\qqq
Both are thus decorrelated in time, a fact that leads to an exact
solution for the correlation functions of the scalar.
The forcing covariance
\s$\CC\s$ is assumed to be a real, smooth, positive-definite function
with rapid decay at spatial infinity so that the forcing
takes place on the (``integral'') scale \s$L\s$.
\vs 0.2cm

The velocity covariance
\s$D\s$ is taken to mimic the situation in real turbulent
flow with structure function
$<(\Nv(t,\Nx)-\Nv(t,0))^2>$ proportional to
\s$\vert\Nx\vert^{\kappa}\s$
for $\kappa>0$.
Concretely, we set
\qq
D^{ij}(\Nx)\s=\s D_0\m\int\ee^{\m i\m\Nk\cdot\Nx}
\s\s(\Nk^2+m^2)^{-(3+\kappa)/2}\s\s(\delta^{ij}-k^ik^j/\Nk^2)
\s\s{_{d^3{\bf k}}\over^{(2\pi)^3}}\ ,
\label{43}
\qqq
where the transverse projector in the Fourier space
ensures the incompressibility of \s$\Nv\s$.
Small $m^2$ is  an infrared cutoff
making the integral convergent for \s$0<\kappa<2\s$.
Writing \s$D(\Nx)\s = \s D(0)-\tilde D(\Nx)\s,$ \s we have
\qq
D^{ij}(0)\s=\s{_{\Gamma(\kappa/2)}\over^{12\pi^{3/2}
\s\Gamma((\kappa-3)/2)}}\s\m\delta^{ij}\s\m D_0\s\m m^{-\kappa}\ ,
\label{div}
\qqq
i.\s e. it diverges with \s$m\to 0\s$, but
the velocity structure function
has a limit
\qq
\lim_{m\to 0}\s\s\tilde D^{ij}(\Nx)\s=\s
D_1\s\left((2+\kappa)\s\delta^{ij}\m
\vert\Nx\vert^{\kappa}\s-\s\kappa\s\m x^ix^j\m
\vert\Nx\vert^{\kappa-2}\right)
\label{td}
\qqq
which is a homogeneous function of \s$\Nx\s$.
\s$D_1\s\equiv\s
{{\Gamma((2-\kappa)/2)}\over{2^{2+\kappa}\pi^{3/2}
\kappa(3+\kappa)\s\Gamma((3+\kappa)/2)}}\s\m D_0\s$ and both
constants have dimension \s${length^{2-\kappa}
\over time}\s$.

\VS{3mm}

We would like to study the statistical properties
of the solutions of Eq.\s\s(\ref{PS}) in the regime
of small \s$\nu\s$, small \s$m\s$ (which may be viewed as
the inverse of another integral scale) and large \s$L\s$.
In particular,
the universality question for the passive scalar
may be formulated as
inquiring about the existence of the limit
of the correlation functions
$<\prod\limits_{n} T(t_n,\Nx_n)>$
in a stationary state of the system when
\s$\nu,\m m,\m L^{-1}\m\to 0\s$ and about the independence
of such a limit of the shape of the source covariance \s$\CC\s$.
We will show that the model possesses an ``inertial'' range
of scales $(\nu/D_1)^{1/\kappa}\ll |x|\ll\min(L,m^{-1})$ where
these correlators become independent of $\nu$,
have a limit as $\nu\rightarrow 0$ and $m\rightarrow 0$
(independent of the order), but in general have {\it non-universal}
(i.e. dependent on the forcing covariance) contributions
involving positive powers of \s$L\s$. \s In
particular we show that the structure functions
\qq
S_{2M}(x)\s\equiv\s <(T(x)-T(0))^{2M}>\s\s\sim\m\s \gamma_{2M}\s
({L/{|x|}})^{\rho_{2M}}\s\m|x|^{(2-\kappa)M}
\label{sf}
\qqq
for $|x|\ll L$
in the $\nu=0$ limit. The amplitudes \s$\gamma_{2M}\s$
are \s$\kappa$- and \s$\CC$-dependent and
the anomalous exponents $\rho_{2M}$ depend on
$\kappa$ but not on $\CC$. We find that \s $\rho_2=0\m$ but
\qq
\rho_4={_4\over^5}\kappa+{\cal O}(\kappa^2)
\label{expo}
\qqq
for small $\kappa$. The H\"older inequality implies
that \s$\rho_{N}\s$ is a convex function of \s$N\s$.
It follows that all \s$\rho_{2M}\s$ for \s$M=2,3,\dots\s$
are strictly positive (\s$\rho_{2M}\geq(M-1)\rho_4\s$)
and that they increase with \s$M\s$.
Thus structure
functions of order four and higher exhibit anomalous scaling
and have explicit integral scale dependence. While we are
able to calculate \s$\rho_4\s$ only for
small $\kappa$ there is no reason
to doubt the generality of the phenomenon
for all $0<\kappa<2$. We want to stress that the non-universality
is due to the pumping scale, not to the dissipation scale,
although the derivatives of the field $T$ have
short distance (non-anomalous)
singularities that are easy to analyze \cite{Russ}
\cite{Proc}.

\VS{3mm}

It is quite well known that in the stationary state the scalar
correlations satisfy linear PDE's \cite{Kraich, Frischpr}.
In the presence
of the UV and IR cutoffs $\nu$ and $m,L$ they have well defined
representations in terms of the Green functions of the corresponding
differential operators which we now recall (for more details see
\cite{my, Sch}).

Suppressing the spatial variable,
the solution of Eq.\s\s(\ref{PS}) with
the initial condition \s$T_0\s$ at \s$t=t_0\s$
takes the form
\qq
T(t)\s=\s R(t,t_0)\s\m T_0\s
\m+\s\int_{t_0}^t\hs{-0.15cm} R(t,s)\s
f(s)\s ds\ ,
\label{FDS}
\qqq
where \s$R(t,t_0)\s$ is given by
the time ordered exponential (\s$t\geq t_0\s$)
\qq
R(t,t_0)\ =\ \CT\ \ee^{
\int_{t_0}^t(\m\nu\m\Delta\m+\m{\bf v}(\tau)\cdot\nabla\m)\s\m d\tau\s}.
\qqq
Thus, to calculate the correlations of $T$ we need to evaluate
expectations
of products of matrix elements of \s$R(t,t_0)\s$.
\s We shall use the tensor
product notation \s$R(t,t_0)^{\otimes\m N}\s$
as a bookkeeping device for all such products.
One then calculates
\qq
< R(t,t_0)^{\otimes N}>
\ =\ \ee^{\m-\m(t-t_0)\s\CM_N}\ ,
\label{Prop2}
\qqq
where $\s\CM_N$ is the differential operator
\qq
\CM_N\s&=\s&-\sum\limits_{n=1}^N
(\nu\s\Delta_{\Nx_n}+\hf\m\CD^{ij}(0)\s\da_{x_n^i}
\da_{x_n^j})\s-\s\sum\limits_{n<n'}\CD^{ij}(\Nx_n-\Nx_{n'})
\s\da_{x_n^i}\da_{x_{n'}^j}\  .
\label{cmn}
\qqq
The gaussian integral of the time ordered exponentials
is calculable due to the independence of $\bf v$'s at different times.
The $\hf\m\CD^{ij}(0)$ term is the contribution of contractions
within a single $R$ and the last terms come from contractions
between different $R$'s. The former one is an ``eddy-diffusivity''
contribution:
since \s\s$\CD^{ij}(0)\m=\m{_1\over^3}\s
\delta^{ij}\s\m\CD^{ll}(0)\s$ and
\s$\CM_1=\s-
(\nu+{_1\over^6}\m\CD^{ll}(0))\m\s\Delta\s,$ \s
it follows from Eq.\s\s(\ref{Prop2}) with \s$N=1\s$
that, in the absence of the sources,
the expectation value of the scalar diffuses
with the effective diffusion constant
composed of the molecular diffusivity \s$\nu\s$
and the eddy diffusivity
\s${1\over6}\m\CD^{ll}(0)\s$. \s
For small \s$m\s$, \s the eddy diffusivity dominates
and the diffusion is driven by the large distance scales
(recall from Eq.\s\s(\ref{div})
that \s$\CD^{ll}(0)=\CO(D_0\s m^{-\kappa})\s$)\m.
\vs 0.3cm

To get hold of the steady state of the scalar, let us first consider
the 2-point function. From (\ref{FDS}), we obtain
\qq
< T(t)^{\otimes\m{2}}>\s&=&\s\ee^{\m-\m(t-t_0)\s\CM_2}\s\s
T(t_0)^{\otimes\m 2}\s+\s\smallint_{t_0}^t ds\s\s
\ee^{\m-\m(t-s)\s\CM_2}\s\s\CC_L\ .
\label{2point}
\qqq
When \s$t_0\rightarrow-\infty\s$, \s the term with
\s$T(t_0)\s$ disappears due to the positivity of
\s$\CM_2\s$ and we obtain
for the steady state
\qq
< T^{\otimes\m{2}}>\ =\ \CM_2^{\s-1}\s\CC_L\ .
\label{2pst}
\qqq
Due to the translational
invariance of $\cal C$ the eddy diffusivity will not
contribute to (\ref{2pst}). Indeed,
note that \s$\CM_2\s$ commutes with (three-dimensional)
translations and in the action on translation-invariant
functions of \s$\Nx_1-\Nx_2\equiv\Nx\s$
reduces to
\qq
\CM_2\s=\s-2\m\nu\s\Delta\s-\s\tilde\CD^{ij}(\Nx)
\s\da_{i}\da_{j}\ .
\qqq
Since \s$\tilde\CD^{ij}(\Nx)\equiv\CD^{ij}(0)-\CD^{ij}(\Nx)\s$
has an \s$m\to 0\s$ limit given by (\ref{td}), so does the operator
\s$\CM_2\s$ in the action on translation-invariant
functions and when \s$\nu\to 0\s$, it becomes
a singular elliptic operator
\s$
\CM_2^{\rm sc}\s=\s -\m D_1\left((2+\kappa)
\s\delta^{ij}\s|x|^{\kappa}\s-
\s\kappa
\s x^i\m x^j\s|x|^{\kappa-2}\right)\da_{i}\da_{j}\s$. \s It
is now easy to analyze (\ref{2pst}) as the various cutoffs $\nu\
 ,\ m\ ,\ L$ are removed using the rotational invariance
of \s$\CM_2\s$. \s In the \s$m\to 0\s$ and \s$\nu\to 0\s$
limits (which commute, we could also take \s$m\s$ proportional
to \s$L^{-1}\s$ with no loss), one obtains for the
two point function \s$F_2(\vert{\bf x}\vert)\m\equiv\s<T({\bf x})\m
T(0)>\s$
\qq
F_2(r)\s=\s\gamma_2\s L^{2-\kappa}\s-\s
{_{2\s\epsilon}\over^{3\m(2-\kappa)\m D_1}}
\  r^{2-\kappa}(1+\CO({_r\over^L}))\ ,
\label{2pnt}
\qqq
where $\gamma_2$ is a non-universal (i.\s\s e. $\CC$-dependent)
constant and $\epsilon =\ha\CC(0)\s$ is
the energy dissipation rate of
the scalar. Note that the non-universal term (a constant)
is annihilated by \s$\CM_2^{\m\rm sc}\s$. This has to be so
if the equation \s$\CM_2^{\m\rm sc}\m F_2=\CC_L\s$ is to be satisfied: the
right hand side becomes universal in the limit \s$L\to\infty\s$
so all non-universal terms in \s$F_2(r)\s$ surviving
in this limit have to be annihilated by \s$\CM_2^{\m\rm sc}\s$.
We shall see this general mechanism limiting possible
non-universal terms also for the higher point functions.
The constant term of \s$F_2\s$
drops out from the $2^{\m\rm nd}$ structure function
which has a universal $L\rightarrow\infty$ limit
so that the exponent $\rho_2=0$. The same universal
result holds approximately in
the whole inertial range \s$\eta\ll r\ll \min(L,m^{-1})\s$, \s
where the Kolmogorov scale \s$\eta=(\nu/D_1)^{1/\kappa}\s$.
\vs 0.4cm

Let us now analyze the higher point correlators. Proceeding as with
the 2-point function,
the steady state solution in terms of the operators ${\cal M}_N$
follows after some simple algebra. For the 4-point function one gets
\qq
<\prod\limits_{n=1}^4T(\Nx_n)>\s=\s
F_4(\Nx_1,\Nx_2,\Nx_3,\Nx_4)\s+\s
F_4(\Nx_1,\Nx_3,\Nx_2,\Nx_4)\s+\s
F_4(\Nx_1,\Nx_4,\Nx_2,\Nx_3)\hs{0.7cm}
\label{481}
\qqq
with the single channel function
\qq
\cr
F_4&=&\CM_4^{-1}\ (\m\CM_2^{-1}\otimes 1\s
+\s 1\otimes\CM_2^{-1})
\ \CC_L\otimes\CC_L\ .
\label{482}
\qqq
Similarly, the higher stationary state
equal-time correlation
functions
$< T^{\otimes\m{2M}}>$ are obtained by symmetrizing
the expressions
\qq
\m\CM_{2M}^{\ \m-1}
(\m\CM_{2M-2}^{\ \m-1}
\otimes\m 1_2\m)\ \cdots\
(\m\CM_2^{\s-1}\otimes\m 1_{2M-2}\m)\m\s
{\CC_L}^{\otimes\m M}
\label{1chan}
\qqq
in their arguments (the odd-point functions vanish).
Expressions (\ref{482}) and (\ref{1chan}) are well
defined for $\nu,m,L^{-1}$ nonzero and we
need to discuss their limits
as these cutoffs are removed.
\vs 0.2cm

The main points of this analysis
are the following. Acting on translation
invariant functions as in (\ref{1chan}), $\CM_N$ becomes
\qq
\CM_N\s=\s-\nu\sum\limits_{n=1}^N\Delta_{\Nx_n}\s+\s
\sum\limits_{n<n'}\tilde\CD^{ij}(\Nx_n-\Nx_{n'})\s\da_{x_n^i}
\da_{x_{n'}^j}
\label{448}
\qqq
i.e. the eddy diffusivity cancels and (\ref{448})
has an \s$m\to0\s$ and \s$\nu\to0\s$ limit
which is a singular elliptic operator \s$\CM_N^{\rm sc}\s$.
It can be shown (\cite{my}) that the Green functions
occurring in (\ref{1chan}) also have limits that are
well defined in the UV and render (\ref{1chan}) finite
for $L$ finite. Thus we need to find the leading
behavior of (\ref{1chan}) as $L\rightarrow\infty$ with
${\cal M}_{2M}$ replaced by \s$\CM_{2M}^{\rm sc}\s$.
\vs 0.2cm

Let us look in more detail at the 4-point function.
Recalling that \s$F_2=\CM_2^{-1}\m\CC_L\s$,
\s it is convenient to view Eq.\s\s(\ref{482})
as a differential equation for \s$F_4\s\m$
that becomes for the connected part \s$F^c_4\equiv
F_4-F_2\otimes F_2\s$
\qq
\CM_4\s\s F^c_4\s=\s\CL\s\m(\m F_2\otimes F_2\m)\ ,
\label{esseq}
\qqq
where \s$\CL\s$ is given by
the sum in (\ref{448}) with $n=1,2$ and
$n'=3,4$. By (\ref{2pnt}), the RHS
of (\ref{esseq}) has a well defined limit
as \s$L\to \infty\s$ given by
\qq
{_{\epsilon^2}\over^{9\s(2-\kappa)^2\s D_1^2}}
\s\s\CL\s\s\vert\NX\vert^{2-\kappa}
\m\vert\NZ\vert^{2-\kappa}
\label{limi}
\qqq
and is a homogeneous
(rotationally-invariant)
function of \s$\NX\equiv {\bf x}_1-{\bf x}_2,\m\NY\equiv
{\bf x}_2-{\bf x}_3\m$ and $\m\NZ\equiv {\bf x}_3-{\bf x}_4\s$
of degree \s$2-\kappa\s$.
It is not difficult to write down a solution
\s${F^c_4}^{\s\rm sc}\s$
of Eq.\s\s(\ref{esseq})
for the limiting case with $\CM_4^{\rm sc}$ and (\ref{limi}) on the
RHS. One easily checks that
\qq
F_4^{\m c\s\rm sc}\ =\ {_{\epsilon^2}
\over^{6\m(2-\kappa)^2(5-\kappa)\m D_1^2}}
\s\s(\s\vert\NX\vert^{2(2-\kappa)}
+\m\vert\NZ\vert^{2(2-\kappa)}\m)
\s-\s{_{\epsilon^2}\over^{9\s(2-\kappa)^2\s D_1^2}}
\s\s\vert\NX\vert^{2-\kappa}
\m\vert\NZ\vert^{2-\kappa}\ ,\
\label{4scal}
\qqq
is such a solution with the use of the decomposition
\m\s$\CM_4^{\rm sc}\s=
\s\CM_2^{\rm sc}\otimes1\s+\s1\otimes\CM_2^{\rm sc}
\s-\s\CL\s$
since \s\s$\CM_2^{\rm sc}\otimes1+1\otimes\CM_2^{\rm sc}\s\s$
vanishes in the action on \s$F_4^{\m c\s\rm sc}\s$
and \s\m$\CL\m\s$ annihilates
functions depending only on \s$\NX\s$ or only on \s$\NZ\s$.
Thus, we deduce that
$\CM_4^{\rm sc}\s\s (F_4^{\m c}-F_4^{\m c\s\rm sc}\m)
\rightarrow 0$ as
$L\rightarrow \infty$. By scale invariance it is thus
reasonable to conclude that the
solution for finite but large \s$L\s$ should differ
from the universal scaling form by zero modes of
\s$\CM_4^{\rm sc}\s$ so that
\qq
F^c_4\ \s-\sum\limits_{0\leq\rho_{4,n}\leq 2(2-\kappa)}
\hs{-0.25cm}L^{\rho_{4,n}}\sum\limits_m\gamma_{nm}\s\s F^c_{4,nm}
\ \ \
\smash{\mathop{\longrightarrow}
\limits_{L\to\infty}}\ \ \ {F_4}^{\s c\s\rm sc}
\label{AsymP}
\qqq
where \s$F^{\m c}_{4,nm}\s$ are homogeneous zero modes
of \s$\CM_4^{\rm sc}\s$ of degree \s$2(2-\kappa)-\rho_{4,n}\s$
and the non-universal coefficients \s$\gamma_{nm}\s$
depend on the source covariance \s$\CC\s$.
\vs 0.2cm

In fact using spectral analysis of ${\cal M}_4$ \cite{my},
(\ref{AsymP}) can be made rigorous (possibly
with logarithmic corrections in $L$ for special values
of $\kappa$). Similar
analysis can be repeated for $N$-point correlators:
non-universal $L$-dependent terms proportional to
homogeneous zero modes of ${\cal M}_{N}^{sc}$ can be present
in the large $L$ asymptotics. We thus face the problem of
finding such zero modes,  of determining whether they are
present in the $N$-point function of $T$ and finally
of finding whether they contribute to the structure
function $S_N$. While such an analysis still eludes us for
general values of $\kappa$, we will now show that
at least for small $\kappa$ the zero modes are present and
dominate the structure functions.

\VS{3mm}

The stationary state correlation functions of the scalar,
given in terms of (\ref{1chan}), are in general non-gaussian,
but they become gaussian as $\kappa\rightarrow 0\m$. \m To
see this, note that in this limit
$\tilde D^{ij}\s=\s 2\s D_1\s\m\delta^{ij}\
$
(having finite \s$D_1\s$ requires
the vanishing of \s$D_0\s$ as \s$\kappa\to 0\s$ in order
to renormalize the ultraviolet divergence
in
(\ref{43})\m; \s$D_0\s$ will never show up below.)
We immediately obtain for the \s$\kappa=0\s$ operators\s:
\qq
\CM_{2,0}^{\rm sc}&=&2\s D_1\s\nabla_{\Nx_1}\cdot\nabla
_{\Nx_2}\s=\s-\m 2\s D_1\s\Delta_\NX\ ,\label{47}\\\cr
\CM_{4,0}^{\rm sc}&=&2\s D_1\sum\limits_{1\leq
n<n'\leq 4}\nabla_{\Nx_n}
\cdot\nabla_{\Nx_{n'}}\cr\cr
&=&-2\s D_1\m\left(\Delta_\NX\s+\s\Delta_\NY
\s+\s\Delta_\NZ\s-\s\nabla_\NX\cdot\nabla_\NY
\s-\s\nabla_\NY\cdot\nabla_\NZ\right)
\label{471}
\qqq
in the difference variables \s\s$\NX\m,\
\NY\m,\m\ \NZ\s$ and using the subscript to refer to $\kappa=0$.
Some straightforward algebra \cite{my}  shows that the expression
(\ref{1chan}) reduces (when symmetrized) to the standard gaussian
expression of sums of product of 2-point functions.

\VS{5mm}

Our strategy is now the following. We shall find
the homogeneous zero modes of
the operator \s$\CM_{4}^{\rm sc}\s$ in perturbation
expansion in powers of \s$\kappa\s$.
\s Eq.\s\s(\ref{td}) implies that (for \s$m=0\s$)
$\tilde D^{ij}(\Nx)\s=\s 2\m D_1(\s\delta^{ij}\s+\s
\kappa\s \s\m R^{ij}(\Nx))\s
+\s\CO(\kappa^2)\ \m\s$ with
\qq
R^{ij}(\Nx)\s=
\m
\delta^{ij}(\ha+\s\ln\vert\Nx\vert\s)\s-\s\ha\s x^i\m
x^j\s\vert\Nx\vert^{-2}
\qqq
Hence, to the first order in \s$\kappa\s$,
$\CM_4^{\rm sc}=
\CM^{\rm sc}_{4,0}\s+\s2\m\kappa\s D_1\s\s V_4$ ,
with
\qq
V_4&=&
-
R^{ij}(\NX)\s\m\da_{X^i}\s\da_{X^j}
\s-\s R^{ij}(\NY)\m\s\da_{Y^i}\s\da_{Y^j}
\s-\s R^{ij}(\NZ)\m\s\da_{Z^i}\s\da_{Z^j}\ \s\cr
&&-\left(R^{ij}(\NX+\NY)-
R^{ij}(\NX)-R^{ij}(\NY)\right)\m\da_{X^i}\s\da_{Y^j}\ \s\cr
&&-\left(R^{ij}(\NY+\NZ)-R^{ij}(\NY)-R^{ij}(\NZ)
\right)\m\da_{Y^i}\s\da_{Z^j}\ \s\cr
&&-\left(R^{ij}(\NX+\NY+\NZ)-R^{ij}(\NX+\NY)
-R^{ij}(\NY+\NZ)+R^{ij}(\NY)\right)
\s\da_{X^i}\s\da_{Z^j}
\qqq

Since \s$\CM_4^{\rm sc}\s$ commutes with $3$-dimensional translations
and rotations and with the permutations of four points
we shall search for its zero modes respecting these symmetries.
Consider first the unperturbed zero modes. The
symmetric zero modes of the lowest
homogeneity of \s$\CM_{4,0}^{\rm sc}\s$
occur in degree zero (constants) and in degree \s$4\s$.
The latter, when expressed in terms of point-differences,
have the form
\qq
&&a\hs{-0.2cm}\sum\limits_{\{n,n'\}}\hs{-0.1cm}
(\Nx_n\hs{-0.05cm}-\hs{-0.05cm}\Nx_{n'})^4\s+\s b
\hs{-0.85cm}\sum\limits_{\{\{n,m\},\m\{n,m'\}\}}\hs{-0.8cm}
(\Nx_n\hs{-0.05cm}-\hs{-0.05cm}\Nx_{m})^2\m
(\Nx_n\hs{-0.05cm}-\hs{-0.05cm}\Nx_{m'})^2
%\ \s\s\cr
\s+\s c\hs{-0.85cm}
\sum\limits_{\{\{n,n'\},\{m,m'\}\}}\hs{-0.8cm}
(\Nx_n\hs{-0.05cm}-\hs{-0.05cm}\Nx_{n'})^2\m
(\Nx_m\hs{-0.05cm}-\hs{-0.05cm}\Nx_{m'})^2\cr
&&\equiv \s a\m F_1\m +\m b\m F_2\m+\m c\m F_3\s\m,
\label{zm2}
\qqq
where the pairs \s$\{n,n'\}\s$ and \s$\{m,m'\}\s$
are assumed different, as well as the pairs \s$\{n,m\}\s$
and \s$\{n,m'\}\s$ and where
$10a+14b+3c=0$ .
\vs 0.2cm

The constant survives as the
eigenvalue of \s$\CM_4^{\rm sc}\s$
for \s$\kappa\not=0\s$. \s
Thus we need to calculate in degenerate perturbation theory how
the fourth degree zero modes change with $\kappa$. For this we
write
$
\CM_{4,0}^{\rm sc}\s=\s-2\m D_1\s(\m\Delta_{\tilde\NX}+\Delta
_{\tilde\NY}+\Delta_{\tilde\NZ})$, where
$\tilde\NX\s=\s\NX\s,\ \ \ \tilde\NY\s=\s
\sqrt{2}(\NY+\hf\NX+\hf\NZ)\s,\ \ \ \tilde\NZ\s=\s\NZ$ .
Denoting \s$R\s\equiv\s(\tilde\NX^2+\tilde\NY^2
+\tilde\NZ^2)^{1/2}=(\hf\hs{-0.16cm}\sum\limits_{\{n,n'\}}
(\Nx_n-\Nx_{n'})^2)^{1/2}\s$,
\s we obtain
\qq
\CM^{\rm sc}_{4,0}\s=\s-\m{_{2\m D_1}\over^{R^8}}\s\da_R\s
R^8\s\da_R\s-\s{_{2\m D_1}\over^{R^2}}\s\Phi
\label{formo}
\qqq
where \s$\Phi\s$ is the Laplacian on the sphere \s$S^8\s$
in the space
of \s$(\tilde\NX,\tilde\NY,\m\tilde\NZ)\s$.
Now pick two linearly independent zero
modes \s$R^4\m f_i\s$, \s$i=1,2\s$, \s of the form
(\ref{zm2}) and look for a homogeneous
zero mode
\qq
R^{4+\kappa\lambda}\s(\m a_1\s f_1\s+\s
a_2\s f_2\s+\s\kappa\s f_3\m)
\label{mode}
\qqq
with a homogeneous degree zero
function \s$f_3\s$ orthogonal
to \s$f_{1,2}\s$ in \s$L^2(S^{8})\s$ \s. We obtain
in the linear order in \s$\kappa\s$
\qq
\CM_{4,0}^{\rm sc}\m
\left(\lambda\s R^4\s\ln{R}\s\m(a_1 f_1+a_2 f_2)
\s+\s R^4\s\m f_3\m\right)\s\m
+\m\s 2\m D_1\s\m V_4\s\m R^4\s(a_1 f_1+a_2 f_2)\s=\s0
\qqq
or, using the form (\ref{formo}) of \s$\CM_{4,0}^{\rm sc}\s$,
\qq
-15\s\lambda\m\s(a_1 f_1+a_2 f_2)\s-\s44\s f_3\s-\s\Phi\s f_3
\s+\s{_1\over^{R^2}}\s\m V_4\s R^4\s(a_1 f_1+a_2 f_2)\s=\s0\ .
\label{pe0}
\qqq
Upon taking the \s$L^2(S^8)\s$
scalar products with \s$f_{1,2}\s$, \s$f_3\s$ drops out
resulting in the relation
\qq
-15\s\lambda\s\sum\limits_{j=1,2}(\m f_i\m,\m f_j)\s\m a_j
\s+\s\sum\limits_{j=1,2}
(\m f_i\m,\s{_1\over^{R^2}}\s V_4\s R^4\s f_j)
\s\m a_j\s=\s0
\label{pe}
\qqq
Hence \s$\lambda\s$ has to solve the equation
\qq
\det\left(\s(\m f_i\m,\s{_1\over^{15\s R^2}}
\s V_4\s R^4\s f_j)\s-
\s\lambda\s\m(\m f_i\m,\s f_j)\right)\s=\ 0\ .
\label{dEt}
\qqq
For the explicit calculation we took $R^4f_1=3F_1-10F_3\m$
and $R^4f_2=-7F_1+5F_2$.
The integrals over the 8-dimensional sphere
are most conveniently done by
using homogeneity to transform them to gaussian integrals
over $\NR^9$. The integration
is straightforward with Maple\footnote{we thank Ezra Getzler
for writing the program for us} and the
matrix in (\ref{dEt}) becomes proportional to
$\left(\matrix{-52-25\lambda & 15+15\lambda\cr
18+15\lambda & -20-20\lambda\cr}\right)$
with the eigenvalues $\lambda_1=-{14/5}$ and $\lambda_2=-1$.
The corresponding eigenfunctions (\ref{mode})
are given by
\s$F^{\m c}_{4,1}=R^{-(2+4/5)\kappa+\CO(\kappa^2)}
(F_1-2F_2+6F_3+\CO (\kappa^2))\s$
and \s$F^{\m c}_{4,2}=R^{-\kappa+\CO(\kappa^2)}
(-7F_1+5F_2+\CO (\kappa))\s$.
\s For large \s$L\s$ the connected 4-point
function takes the form
\qq
<\prod\limits_{n=1}^4 T({\bf x}_n)>^c
&=&\gamma_{4,0}\s L^{4-2\kappa}
\s+\s\gamma_{4,1}\s (L/R)^{4\kappa/5+\CO(\kappa^2)}\s
R^{-2\kappa}(F_1-2F_2+6F_3+\CO(\kappa))\ \s\cr
&+&\gamma_{4,2}\s (L/R)^{-\kappa+\CO(\kappa^2)}\s
R^{-2\kappa}(-7F_1+5F_2+\CO(\kappa))\cr
&+&<\prod\limits_{n=1}^4 T({\bf x}_n)>^{c\s{\rm sc}}\s
+\s\s\CO((L/R)^{-2+\CO(\kappa)})
\qqq
uniformly in small \s$\kappa\s$. \s Since the connected
correlation vanishes and  \s$
<\prod\limits_{n=1}^4 T({\bf x}_n)>^{c\s{\rm sc}}\s$
reduces to \s${\epsilon^2\over
360\m D_1^2}\m(3F_1-10F_3)\s$  for \s$\kappa=0\s$,
\s we infer that \s$\gamma_{4,0}=\CO(\kappa)\s$,
\s$\gamma_{4,1}={\epsilon^2\over 216\m D_1^2}+\CO(\kappa)\s$
and \s$\gamma_{4,2}={\epsilon^2\over 540\m D_1^2}+\CO(\kappa)\s$.
The result (\ref{sf}) for \s$N=2\s$
follows with \s$\gamma_4={\epsilon^2\over 3\m D_1^2}+\CO(\kappa)\s$
and \s$\rho_4\s$ given by (\ref{expo})
since only the \s$F^{\m c}_{4,1}\s$-term gives
non-zero contribution to the structure function.
The zero mode, with $\m\lambda=-1\m$, \m is actually
obtained from a zero mode of ${\cal M}_3$ by extending it to
a function of four ${\bf x}_i$'s by symmetrizing. This is a
general feature: zero modes of ${\cal M}_N$ give rise to
zero modes of ${\cal M}_{2M}$ for $2M>N$. These however do not
contribute to the structure functions $S_{2M}$. The only
zero mode of ${\cal M}_{2M}$ that contributes, is the unique
one not coming from the lower dimensional ${\cal M}_N$'s,
namely the one that at $\kappa=0$ is gotten from the
monomial $\prod_{i=1}^{M}(x_{2i-1}-x_{2i})^2$ by symmetrizing
and subtracting partial traces.
It gives rise to the dominant
contribution to \s$S_{2M}\s$ which has to be present
by the H\"{o}lder inequality.
\vs 0.3cm

The asymptotic behavior of the scalar correlation functions
encodes a subtle information about the behavior
of the Green functions of the singular multibody operators
\s$\CM_{N}\s$ with continuous spectrum. The reduction
of its study to that of discrete spectrum
operators given by \s$\CM_{N}$'s \m acting on homogeneous
functions should be thought of as realizing
a renormalization group
type approach to the model, with the homogeneous
zero modes of \s$\CM_N\s$ playing the role of
relevant interactions. This may be the most important
hint from the above exact solution for the anomalous scaling
of the passive scalar.

\VS{5mm}

\noindent {\it Acknowledgements}.
We would like to thank the Mittag-Leffler Institute,
where this work was started, for hospitality.
Discussions with Uriel Frisch,
Robbert Kraichnan, Itamar Procaccia and Achim Wirth
are acknowledged.
\vs 1cm

\end{document}